\documentclass[aps,prb,twocolumn,superscriptaddress,showpacs]{revtex4}
\usepackage{graphicx}
\usepackage{dcolumn}
\usepackage{bm}

\begin{document}

\title{Common origin of the two types of magnetic fluctuations in iron chalcogenides}

\author{Songxue~Chi}

\email{chis@ornl.gov}

\altaffiliation{Current address: Oak Ridge National Laboratory, Oak Ridge, Tennessee 37831, USA}

\affiliation{NIST Center for Neutron Research, National Institute of Standards
and Technology, Gaithersburg, Maryland 20899-6102, USA}
\affiliation{Department of Materials Science and Engineering, University of Maryland, College Park, Maryland 20742, USA}

\author{J. A. Rodriguez-Rivera}
\affiliation{NIST Center for Neutron Research, National Institute of Standards
and Technology, Gaithersburg, Maryland 20899-6102, USA}
\affiliation{Department of Materials Science and Engineering, University of Maryland, College Park, Maryland 20742, USA}

\author{J.~W.~Lynn}
\affiliation{NIST Center for Neutron Research, National Institute of Standards
and Technology, Gaithersburg, Maryland 20899-6102, USA}

\author{Chenglin~Zhang}
\affiliation{Department of Physics and Astronomy, The University of Tennessee,
Knoxville, Tennessee 37996-1200, USA}

\author{D. Phelan}
\affiliation{NIST Center for Neutron Research, National Institute of Standards
and Technology, Gaithersburg, Maryland 20899-6102, USA}

\author{D. K. Singh}
\affiliation{NIST Center for Neutron Research, National Institute of Standards
and Technology, Gaithersburg, Maryland 20899-6102, USA}
\affiliation{Department of Materials Science and Engineering, University of Maryland, College Park, Maryland 20742, USA}

\author{R. Paul}
\affiliation{Materials Measurement Laboratory, National Institute of Standards
and Technology, Gaithersburg, Maryland 20899-6102, USA}

\author{Pengcheng~Dai}
\affiliation{Department of Physics and Astronomy, The University of Tennessee,
Knoxville, Tennessee 37996-1200, USA}
\affiliation{Beijing National Laboratory for Condensed Matter Physics, 
Institute of Physics, Chinese Academy of Sciences, Beijing 100080, China}

\date{\today}

\begin{abstract}
We use inelastic neutron scattering to study the low energy spin excitations in moderately doped non-superconducting Fe$_{1.01}$Te$_{0.72}$Se$_{0.28}$. 
The spin excitations in this system contain components near (0.5,0,0) and (0.5,0.5,0) in a-b plane reciprocal lattice units using tetragonal unit cell notation (a=b=3.772 A and c=6.061 A).  At low energies the scattering is centered around (0.5,0,0).  With increasing energy, the spectral weight of low energy spin excitations centered around (0.5,0,0) abruptly shifts around 3 meV to the incommensurate spin excitations centered around (0.5,0.5,0). However both types of spin fluctuations exhibit the identical temperature dependence. These results indicate that the (0.5,0,0) type spin excitations and the incommensurate excitations around the (0.5,0.5,0) position have a common origin and both must be taken into account to understand the nature of magnetism and superconducting pairing in the iron chalcogenides.
\end{abstract}


\maketitle

\begin{center}
${\textbf{I.  INTRODUCTION} }$
\end{center}

The iron chalcogenides Fe$_{1+y}$Te$_{1-x}$Se$_x$ have the $\alpha$-PbO structure, which contains layers of Fe squares with the chalcogen atoms residing alternately above and below the centers of the squares, the same way in which FeAs layers are formed in the iron pnictides \cite{hsu,mhfang}.
The superconductivity arises when sufficient Se replaces Te in the antiferromagnetically ordered parent phase Fe$_{1+y}$Te \cite{mhfang}.
Despite the similarities in crystal structure and Fermi surface topology \cite{Subedi,Xia}, the parent compounds of the iron pnictides and iron chalcogenides have very different magnetic structures. The pnictides have a single stripe in-plane collinear ($C$-type, shown in Fig. 1(a)) antiferromagnetic (AFM) structure \cite {JZhao, Qing} characterized by the wave vector (0.5,0.5,0.5) in the notation of the tetragonal lattice as highlighted by the shaded area in Fig. 1(a), which coincides with the Fermi surface (FS) nesting wave vector \cite{IMazin} and the wave vectors of neutron spin resonance in superconducting samples \cite {Lumsden1,Chi}. The nonsuperconducting Fe$_{1+y}$Te has a diagonal double-stripe bicollinear order ($E$-type, Fig. 1(b)) modulated along the $(0.5,0,0.5)$ direction \cite {Bao,SLi1}, whose in-plane component is 45$^{\circ}$ away from the FS nesting wave vector (0.5,0.5,0) where, curiously enough, the spin resonance of superconducting Fe$_{1+y}$Te$_{1-x}$Se$_{x}$ is found \cite{Mook,Qiu}. The disparity between the static order and Fermi surface nesting as well as the large ordered moment \cite{RongweiHu,Martinelli,SLi1} in iron chalcogenides, have fueled the already heated debate about the nature of magnetism in the iron based superconductors. The controversy over the magnetism in the iron chalcogenides has been mainly centered on whether it originates from itinerant electrons \cite {Han1,Han2,Paul}, localized moments \cite {Ma, Fang, Moon} or both \cite{Arita, Kou, Yin, Johannes, JiangpingHu}. 

As a good probe to magnetism, the spin dynamics in the iron chalcogenides has been extensively studied with inelastic neutron scattering (INS) measurements. In undoped Fe$_{1+y}$Te, the low energy excitations start exactly or closely to the (0.5,0,0) position depending on the amount of excess Fe \cite{Lipscombe, Igor,CStock}. However, this dispersion about the magnetic zone center disappears at higher energies, and new rings of excitations emerge above 60 meV around integer positions such as (1,0,0). The ring centered at (1,0,0) disperses inward and eventually becomes a spot before disappearing above 275 meV \cite{Lipscombe}. In the 27$\%$ Se-doped non-superconducting (NSC) sample \cite{Lumsden2}, the magnetic response around (1,0,0) starts from the lowest measured energy and forms incommensurate magnetic (ICM) quartets instead of a ring, but with increasing energy it evolves into a ring. The dispersion is still steep and persists to energies beyond 250 meV.
The normal state magnetism in superconducting FeTe$_{1-x}$Se$_{x}$ \cite{Lumsden2,Sunghoon,Argyriou,SLi2,Xu1} is very similar to that of the non-superconducting compounds except that the quartets never become a ring at high energies \cite{Lumsden2}. The observed a-b plane spin excitations in Fe$_{1+y}$Te$_{1-x}$Se$_x$ are summarized in Fig. 1(c) as schematics in reciprocal space.

In most Se-doped compounds the (0.5,0,0) excitations are still present at low energies. Their intensity diminishes with increasing Se-doping and becomes very weak in superconducting samples \cite {Lumsden2, TJLiu, Xu1}. It is therefore believed that the (0.5,0,0) type spin correlations have a deleterious effect on the superconducting pairing. Since chemical inhomogeneity, impurity and phase separation exist in these materials \cite{HefeiHu, Johnston, Joseph}, 
it is unclear whether these (0.5,0,0) spin fluctuations are intrinsic to the system or are a result of an undesirable phase or domains segregated from the primary magnetism. 

To address the relationship between these two types of coexisting magnetic fluctuations, we have carried out INS measurements on   moderately doped non-superconducting Fe$_{1.01}$Te$_{0.72}$Se$_{0.28}$. Our results indicate that the (0.5,0,0) type excitations dominate the lowest energy spectral response, but then the strength of the scattering abruptly shifts to the ICM excitations centered about (0.5,0.5,0), while the two types of spin fluctuations exhibit the identical temperature dependence.  Taken together, we conclude that the two types of magnetic excitations have a common origin. 

\begin{center}
${\textbf{II.  EXPERIMENTAL} }$
\end{center}

High quality single crystals of $\alpha$-phase Fe$_{1+y}$Te$_{1-x}$Se$_{x}$  were prepared with nominal composition of $x=0.3$ 
using the flux method and co-aligned with neutrons. 
The actual compositions of the samples were determined with prompt gamma activation analysis (PGAA) on beamline NG-7 at the NIST Center for Neutron Research (NCNR).
Neutron scattering measurements were conducted on the cold neutron triple axis spectrometers (TAS) MACS and SPINS  and the thermal TAS BT-7 at NCNR. Pyrolytic graphite (PG) was used as monochromator and analyzer for all the measurements and as filter for BT-7 measurements. A BeO filter was used on SPINS with fixed final energy $E_f$ of 3.7 meV and horizontal collimations of open-80'-S-80'-open. Double focusing monochromator, Be filter, and all 20 channels of the detection system \cite{Jose} were employed on MACS with $E_f$ fixed at 5 meV. Fixed $E_f$ of 35 meV and Open-25'-S-25'-120' collimations were used for INS measurements on BT-7 \cite{BT7}. Open-25'-S-25'-50' with $E_f$=14.7 meV was used for elastic scattering on BT-7. 
The momentum transfer $\textbf{Q}$ at $(q_x,q_y,q_z)$ is defined as $(H,K,L)=(q_xa/2\pi,q_yb/2\pi,q_zb/2\pi)$ reciprocal lattice units (r.l.u.) in the tetragonal unit cell ($P4/nmm$ space group). The lattice parameters of the tetragonal unit cell are  $a=b=3.772~ \text{\AA}$ and $c=6.061~ \text{\AA}$ at $T = 1.5$ K. Elastic measurements were taken in both the $(H,K,0)$ and $(H,0,L)$ scattering planes, while the INS measurements were concentrated in the $(H,K,0)$ plane.  
The error bars shown in the figures are statistical in nature and represent one standard deviation.

\begin{center}
${\textbf{III.  RESULTS AND DISCUSSION} }$
\end{center}

\begin{figure}
\includegraphics[width=3.2in]{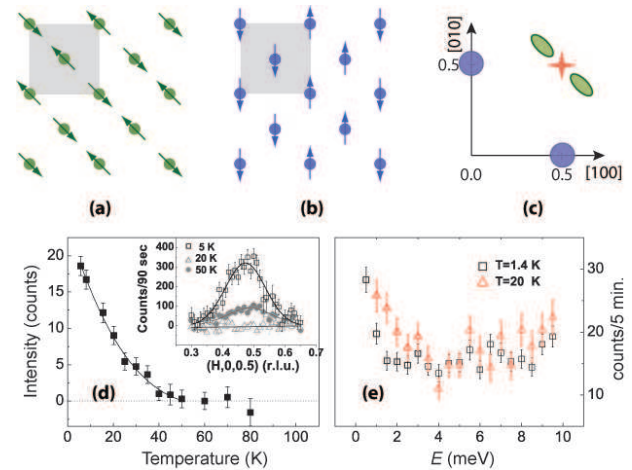}
\caption{\label{fig1} (Color online)  
(a) The schematic C-type collinear AFM order of the Fe moments in the a-b plane for iron pnictides such as LaFeAsO \cite{Jeff}. The shaded area represents the tetragonal unit cell which is used in the study. (b) The ab-plane projection of the E-type bicollinear AFM spin structure for iron telluride FeTe. (c) Schematics of the observed spin excitations for Se-doped FeTe in reciprocal space. The red star shows the position where FS nesting and the spin resonance has been observed for both pnictides and chalcogenides. The solid blue circles represents the low energy spin fluctuations at (0.5,0,0) and the filled green ellipses show the ICM excitations centered around the (0.5,0.5,0) position. (d) Temperature dependence of the integrated intensity of scans along (H,0,0.5) taken on BT-7. The inset shows the background-subtracted H-scans at some typical temperatures. (e) Constant-$\textbf{Q}$ scans at (0.5,0.5,0) measured on SPINS at T=1.4 K and T=20 K. There is no evidence of a spin resonance or the development of a spin gap. 
}
\end{figure} 

We first determine the actual stoichiometry of our Fe$_{1+y}$Te$_{1-x}$Se$_x$ crystals since these can differ 
significantly from the nominal compositions \cite{HefeiHu}.  For this purpose,
prompt gamma activation analysis (PGAA) was carried out on a small piece of single crystal.  
PGAA is a nondestructive technique using neutron adsorption to simultaneously determine the presence and accurate quantities of various elements in a compound \cite{Rick1,Rick2}.  We find that the Fe:Te:Se molar ratio is 1.009:0.721:0.279. Relative expanded uncertainties for PGAA data are estimated at less than 5$\%$.
 
\begin{figure}
\includegraphics[width=3.4 in]{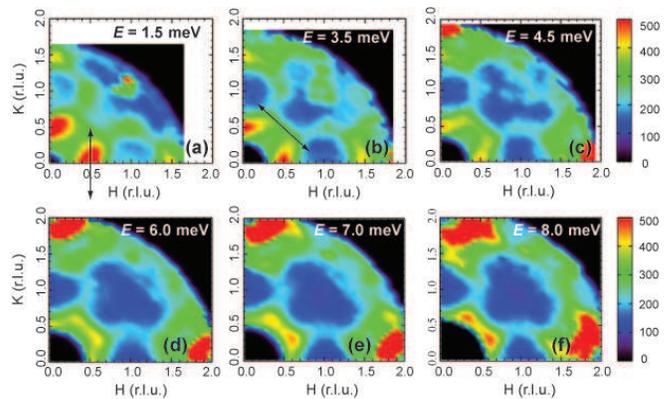}
\caption{\label{fig2} (Color online) Contour plots of neutron scattering intensity in the $(H,K,0)$ plane at energy transfers of (a) 1.5 meV (b) 3.5 meV, (c) 4.5 meV, (d) 6.0 meV (e) 7.0 meV and (f) 8.0 meV. The data were collected at T = 1.5 K on MACS. All the panels are plotted with the same color scale to show the intensity variations. The arrows in (a) and (b) show the directions of constant-\textit{E} cuts plotted in Fig. 3. 
}
\end{figure}

Neutron diffraction measurements were carried out on a small piece of single crystal from the same batch as the crystals used for the inelastic neutron scattering. The crystal was aligned in the $(H,0,L)$ scattering plane. $H$-scans were performed at various temperatures. 
The static magnetic order survives in this compound as short range static order centered at the wave vector (0.47,0,0.5). This is consistent with the previous reports that Se-doped bulk non-superconducting samples have short range order centered at incommensurate wave vector (0.5-$\delta$,0,0.5) \cite{Xu1,Wen,Katayama}. The $\delta$ value, 0.03 in this case, can be tuned by both the Se and Fe concentration \cite {Bao,Wen}. \textit{H}-scans at some typical temperatures are shown in the inset of Fig. 1(d). The magnetic peaks are much broader than the instrumental resolution with Lorentzian fits giving a width of 1.6 (r.l.u.) corresponding to in-plane correlation length of 3.7(8) $\text{\AA}$ at 5 K. This is in agreement with that of similar doping FeTe$_{0.7}$Se$_{0.3}$ \cite {Wen}. The integrated intensity of the $H$-scans at different temperatures is plotted in Fig. 1(d). Upon warming the peak intensity is gradually suppressed without an abrupt transition or change of peak position. The magnetic intensities cannot be detected above 60 K. The diffuse nature of the short-range order is reflected by the concave shape of the intensity-temperature curve.  
Constant-\textit{Q} scans of inelastic neutron scattering at (0.5,0.5,0) at 1.4 K and 20 K are shown in Fig. 1(e). The overall spectrum shows little temperature dependence except some enhancement at 20 K at energies below 4 meV due to thermal population. Neither a spin resonance nor the development of a spin gap at low temperature can be found, confirming the absence of bulk superconductivity.  

The above characterizations place our sample in the intermediate doping part of the phase diagram where the long range AFM order is suppressed but bulk superconductivity has not emerged. In this region, there are reports about weak charge carrier localization \cite{TJLiu} and spin glass ordering \cite{Paulose,Babkevich,Katayama}. The magnetic and superconducting properties are sensitive  not only to the Te/Se ratio, but also to the variation of Fe content \cite{Bao, Wen, Xu1, Bendele, Katayama, Rodriguez, Xu2, CStock, McQueen}.
The modest amount of excess Fe in our sample ensures that no significant complications arise due to the interstitial iron.

Twenty grams of single crystals were co-aligned for the INS measurements which focused on the $(H,K,0)$ scattering plane because of the weak $L$-dependence \cite {Lipscombe,Lumsden2, Argyriou}. For those measurements we drop the L coordinate for simplicity and present the data in terms of $(H,K)$ only. In order to simultaneously monitor the two excitations we need to survey a wide region of momentum space in the low energy transfer range. MACS is ideal for this type of measurement because of the multiple detection systems \cite {Jose}. An empty sample holder in the same sample environment was also measured and used as background subtraction. Fig. 2(a)-(f) show some typical contour plots at 1.5 K using data folded into the first quadrant of the scattering plane. At low energy transfers the (0.5,0) type excitation is dominant with a broad peak. At the equivalent position of higher Brillouin zones, such as (1,0.5) and (1.5,0), peaks are weaker because of the decreased magnetic form factor.  As the energy increases the (0.5,0) type scattering quickly diminishes in intensity while the two ICM peaks around (0.5,0.5) start to appear and intensify. These two peaks are from separate sets of quartets of scattering about (1,0) and (0,1) respectively. We hereby call the excitation represented by these two peaks the ICM excitations to distinguish from the (0.5,0) excitations, and to avoid confusion about the Brillouin zone center. Also visible is the acoustic phonon mode stemming from  (1,1) at low energies and from (2,0)/(0,2) at higher energies.

\begin{figure}
\includegraphics[width=3.2in]{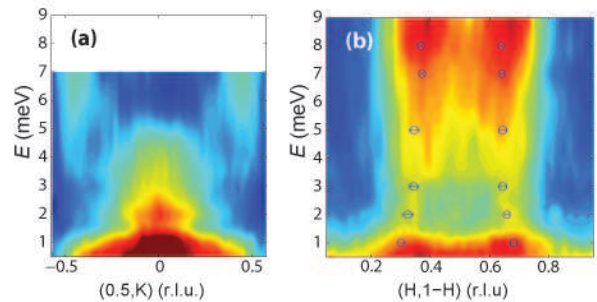}
\caption{\label{fig3} (Color online)  Contour plot of combined cuts for the T=1.5 K data in the (a) $[0.5,K]$ direction and (b) the $[H,1-H]$ direction through (0.5,0.5) as a function of energy. Both figures are plotted on the same energy scale so that the correspondence between the two excitations can be seen. The open circle data in (b) are the result of a two-gaussian fit to the peaks shown in Fig. 5(a), which are measured using the high resolution configuration on SPINS.
}
\end{figure}

\begin{figure}
\includegraphics[width=3.2in]{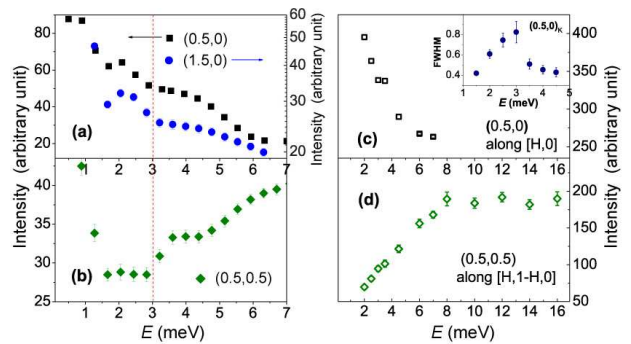}
\caption{\label{fig4} (Color online) (a) Constant-Q cut along \textit{E} at (0.5,0) [black solid squares] and (1,0.5) [blue sold circles]. (b) Cut at (0.5,0.5) along \textit{E}. The shaded area is the energy range where the strong diffuse scattering is present. The dashed line shows roughly where the abrupt changes of intensities occur. (c) Integrated intensity of the cut through (0.5,0) along the H-direction with background subtracted. (d) The sum of fitted areas of the two peaks from the transverse scan through (0.5,0.5). Data obtained from BT-7 are used for \textit{E}$\geq$8 meV.
}
\end{figure}

At various energies, we performed cuts along $[0.5,K]$ through the centers of both (0.5,0) and (0.5,0.5), which enables us to track the evolution of the two excitations simultaneously.  Constant-\textit{E} cuts in the transverse $[H,1-H]$ direction through (0.5,0.5) were also conducted which yield two symmetric ICM peaks that gives the dispersion of the ICM excitation. The combinations of these two cuts as a function of energy are shown in Figs. 3(a) and 3(b). The $[0.5,K]$ cuts at energies above 7 meV are not included because the low-$Q$ space could not be reached with the chosen fixed final energy.
Fig. 3 (a) shows that the disappearance of the broad peak at (0.5,0), as energy increases, is a three-stage process instead of a continuous one. The broad quasi-elastic scattering shows a sharp drop of intensity at about 1.5 meV, followed by another abrupt drop at about 3 meV before the signal disappears above 6 meV. Fig. 3(a) also shows the gradually emerging excitation at (0.5,-0.5) and (0.5,0.5). The complete disappearance of the (0.5,0) excitation at around 6 meV is correlated with the opening of a spin gap for the ICM peaks around (0.5,0.5), as shown in Fig. 3 (b). In this partial gap the ICM peaks around (0.5,0.5) are suppressed in intensity, but their peak profiles remain.
Remarkably, the energies where the twin ICM peaks abruptly change spectral weight coincides with those where the opposite sudden change occurs for the (0.5,0) spectrum. This reciprocal interplay between the two spin excitations can be more easily seen in Fig. 4. The constant-$Q$ scans at these two wave vectors show an opposite energy dependence. The presence of the (0.5,0) fluctuations is compensated by the gap opening in the ICM fluctuations near (0.5,0.5). The dashed lines in Fig. 4(a) indicate the energies at which the two spectra show abrupt changes. The intensity of the constant-$Q$ scan at (1,0.5) is also plotted on a log scale in Fig. 4(a) confirming the energy dependence of the (0.5,0) correlation. 

In order to obtain the integrated intensity for the (0.5,0) spectrum, we performed cuts through (0.5,0) in the $H$-direction, instead of the $K$-direction to avoid the component of the excitation near (0.5,0.5). Because of the unreachable Q-space at higher energies, only half of the thereby obtained peak is integrated and plotted against energy in Fig. 4(c). The integrated intensity of the two ICM peaks around (0.5,0.5) is also plotted in Fig. 4(d). The constant-\textit{E} scans in the transverse direction through (0.5,0.5) were also performed with the thermal triple axis instrument BT-7. The raw data up to \textit{E} = 16 meV and lines of fits to two Gaussians are displayed in Fig. 5(b). The ICM excitation shows very little variation in dispersion and in spectral weight between 8 and 16 meV. The sum of fitted areas of the two Gaussians, together with that of the constant-\textit{E} cuts from the MACS data shown in Fig. 2, is plotted as a function of energy in Fig 4(d). Identical scans at \textit{E} = 8 meV using the two instruments were used to normalize the overall intensities. The spectrum of the ICM excitations about (0.5,0.5) remains constant above 7 meV. This is important because it means that the gap of the ICM excitations below 7 meV is not compensated by spectral gain at higher energies at the same wave vector as is the case in the superconducting compounds. This again indicates that spectral weight is transfered between the two wavevectors and that the (0.5,0) spectrum is at cost of the ICM spectrum. 
Clearly, the spectral weight for these two types of excitations is inversely correlated, which rules out electronic phase separation or magnetic inhomogeneity as the origin of the two types of magnetic correlations \cite{xbhe}.

\begin{figure}
\includegraphics[width=3.2in]{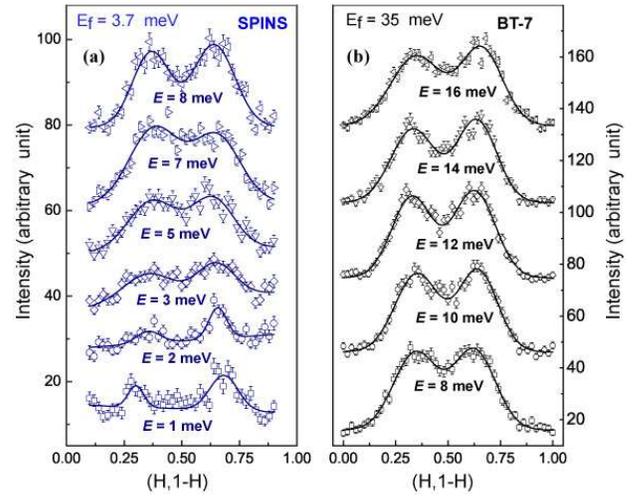}
\caption{\label{fig5} (Color online) Typical scans along the [H,1-H] direction at various energy transfers carried out on (a) SPINS and (b) BT-7 at T = 1.5 K. The lines are Gaussian least squares fits convoluted with instrument resolution. The intensities and fits are shifted up sequently by an equal amount between adjacent scans for clarity. 
}
\end{figure}

Another conspicuous feature in Fig. 3(b) is the dispersion toward (0.5,0.5) below 3 meV before its steep outward dispersion. This hourglass shaped dispersion has been observed in both superconducting and over-doped non-superconducting systems \cite{SLi2, Xu2}. For this under-doped sample the inward dispersion is less pronounced and is more of a bell shape. To confirm this inward dispersion and the intense quasielastic scattering below 1.5 meV, we repeated the $[H,1-H]$ scans on the cold neutron TAS instrument SPINS which, with $E_f$=3.7 meV and horizontally flat monochromator, offers better energy and $Q$ resolution. Fig. 5(a) shows these transverse scans through (0.5,0.5) at various fixed energy transfers. The low background and distinct profile of the two peaks at $E = 1$ meV in Fig. 5(a), in contrast to the broad quasielastic scattering extending to $E=1.5$ meV in Fig. 3(b) and Fig. 4(b), is the result of improved resolution. The peaks are fit with two Gaussians convoluted with the instrumental resolution. The $Q$-positions obtained are over-plotted in Fig. 3(b) with open circles. The results are consistent with the MACS data and the bell shaped dispersion is clearly visible. 
The inward dispersion stops around 3 meV. 
Comparing our data with the studies of other compositions \cite{SLi2, Xu2}, it seems the increasing Se doping pushes the saddle point to higher energies.
It should be also noted that in the energy range where the bell shape of the ICM excitation occurs, there is an abnormal change of the (0.5,0) spin spectrum as shown in Fig. 3(a). The change happens to both the line-width (inset of Fig. 4(c)) and the peak intensity (Fig.4(a)). These anomalies are confirmed by a similar \textit{Q}-\textit{E} plot (not shown) for the cut through (1,0.5), which shows identical anomalies in intensity and linewidth. 

We now turn to the temperature dependence of the spin excitations. The $(H,K,0)$ planar maps have been obtained at different temperatures between 1.5 K and 308 K for four typical energy transfers: 1 meV in the quasielastic region, 4.5 meV in the spin gap, 7.0 meV at the verge of the gap and 10 meV above the gap. Fig. 6 shows the combined $[0.5,K]$ cuts at \textit{E} =1 meV, and the $[H,1-H]$ cuts at \textit{E} = 1 meV and 10 meV as a function of temperature. The integrated intensity of these cuts at all the above-mentioned energies are plotted in Fig. 7 (a) and (b).
Both types of magnetic correlations are so robust that they maintain their well-defined features up to the highest measured temperature for all the energy transfers. On warming the (0.5,0) spectrum at \textit{E} = 1 meV starts to gain intensity at about 60 K, where the static order disappears, reaches its maximum at about 80 K and gradually decreases at higher temperatures. The ICM spectrum around (0.5,0.5), however, is clearly gapped below the transition temperature of the static AFM order, as shown in Fig. 6(b) and Fig. 7(b). As the system is heated from the short-range static ordered phase into the paramagnetic phase the static component also transfers to the background, resulting in the abrupt broadening of the linewidth of the (0.5,0) spectrum, as shown in Fig. 6(c). At \textit{E}=7 meV the intensity of the ICM spectrum is less affected by the static order below 60 K. At \textit{E} = 10 meV (Fig. 6(d)), the ICM intensity remains unaffected by the static order and gradually increases monotonically all the way above 300 K. 

\begin{figure}
\includegraphics[width=3.2in]{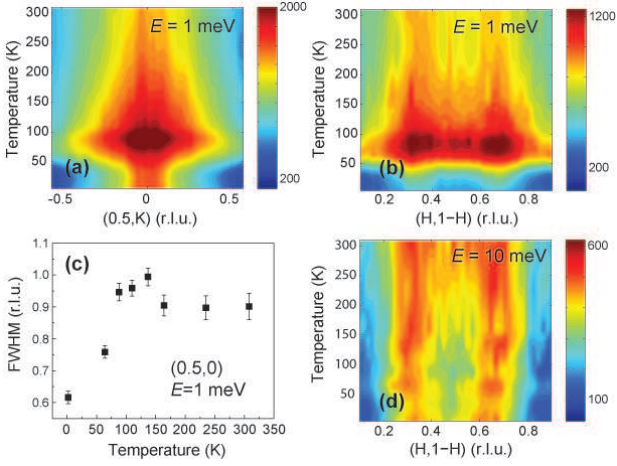}
\caption{\label{Fig6} (Color online) Temperature dependence of the cut (a) along [0.5,K] at \textit{E} = 1 meV and (b) along [H,1-H] at \textit{E} = 1 meV and (d) \textit{E} = 10 meV. Note the change in the vertical color bar scales made to better show the change of intensities. (c) Temperature dependence of the full-width-at-half-maximum (FWHM) of the (0.5,0) spectrum at \textit{E} = 1 meV.
}
\end{figure}

\begin{figure}
\includegraphics[width=3.2in]{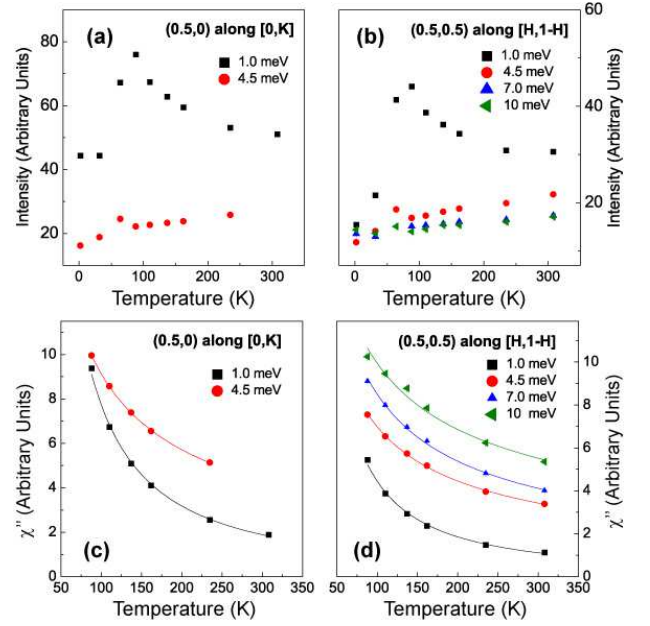}
\caption{\label{fig7} (Color online) Temperature dependence of the integrated intensities of (a) the (0.5,K) scan at \textit{E} = 1 meV and \textit{E} = 4.5 meV, and (b) the $[H,1-H]$ scan through (0.5,0.5) at  \textit{E} = 1 meV, 4.5 meV, 7 meV  and 10 meV. (c) Temperature dependence of the dynamic susceptibility in the paramagnetic phase (60 K$<$T$<$310 K) obtained from the (0.5,0) type of excitations and (d) the (0.5,0.5) ICM excitations.  The lines in (c) and (d) are fits to the power law $\chi''=CT^{\beta}$.
}
\end{figure}                                                                                                                                                                                                               

The neutron scattering intensity $S_T(Q,\omega)$ at temperature T is related to the imaginary part of the dynamic susceptibility $\chi(Q,\omega)_T''$ through $S_T(Q,\omega)=[n(\omega,T)+1]\chi_T''(Q,\omega)$, where $n(\omega,T)$ is the Bose factor. In order to investigate the temperature dependence of the dynamic susceptibility in the paramagnetic phase, the $S_T(\omega)$ intensities for T$>$ 80 K in Fig. 7(a) and (b) are converted to the values of dynamic susceptibility $\chi(\omega)_T''$ and plotted in Fig. 7(c) and (d). The scattering at energy transfers of 7 meV and 10 meV increases with increasing temperature as expected due to the thermal factor, but the susceptibility decreases with T indicating that the intrinsic strength of the magnetic scattering decreases as the scattering evolves to higher temperature. The T-dependencies of the dynamic susceptibility in the paramagnetic phase (T$>$60) were fit to the power law $\chi''=CT^{\beta}$. The $\beta$ values obtained for wavevectors (0.5,0) and (0.5,0.5) at \textit{E}=1 meV, denoted by black squares in Fig. 7(a) and 7(b) (solid and empty), are -1.28$\pm$0.02 and -1.26$\pm$0.03 respectively. Similarly the $\beta$ values at \textit{E}=4.5 meV (red circles) for these two vectors are -0.68$\pm$0.01 and -0.65$\pm$0.01. Note that the dynamic susceptibility of the two excitations have the identical temperature dependence. This is further evidence that both types of excitations have a common origin. At an energy of 7 meV (Fig. 7(b)) where the (0.5,0) spectrum is almost completely depleted, the $\chi''$ for (0.5,0.5) continues to follow the same relationship with $\beta$=-0.66$\pm$0.01. At an energy transfer of 10 meV, $\beta$ becomes -0.54$\pm$0.03.

For the superconductors, the high energy magnetic excitations remain  unchanged when cooled from the normal state to below T$_C$ \cite {Argyriou,Xu2}. It is the low energy part of the magnetic excitation spectrum that responds to the formation of superconducting pairs, as would be expected. These changes include the opening of a spin gap and the development of a spin resonance. In our NSC system, the gap still develops for the ICM spectrum centered around the same (0.5,0.5) position, but without the development of superconductivity or a spin resonance. Instead, we have the (0.5,0) type excitations, which apparently correspond to that part of magnetism that is needed for the spin resonance in the superconducting state, as only the (0.5,0) type excitations are suppressed when the superconductivity and the associated spin resonance develop. Now that we know that the two type of excitations have a common origin, suppressing the (0.5,0) type correlations through doping Se cannot be simply understood as eliminating a coexisting phase. Rather one has to treat the two excitations as one problem when trying to reveal the driving force for magnetism and superconductivity. In that sense, the itinerant electrons alone may not be able to provide a complete answer. 
The inter-band Fermi surface nesting describes the main features of the ICM excitations such as the incommensurate excitations \cite{Argyriou} and  the hour-glass dispersion near (0.5,0.5) \cite {SLi2}, but the Fermi surface near the X point has not been found \cite {Xia,Nakayama} yet to support the nesting scenario for the (0.5,0) excitation \cite{Han1,Han2}. 

In a local moment picture, the magnetic ground state is governed by superexchange interactions. The contest between the collinear and bi-collinear order is controlled by the competition between J$_1$, J$_2$ and J$_3$, which have different chalcogen height dependencies \cite{Ma,Fang,Moon}. As Se replaces Te, the chalcogen height is reduced \cite{Bendele} which results in increased J$_1$, J$_2$, decreased J$_3$, and consequently a less favored bicollinear order at (0.5,0). The spin wave spectrum centered at (0.5,0) that extends up to 60 meV \cite{Lipscombe} in undoped Fe$_{}$Te is suppressed to lower energies, and is completely taken over by the ICM magnetism as the Se content increases well into the superconducting region. This scenario is supported by the reciprocal interplay between the two types of excitations and the temperature dependence of the low-\textit{E} dynamic susceptibility presented in this study. However, it is not possible to understand the low energy features such as the hour-glass dispersion \cite{SLi2,JiangpingHu} and the abnormal change of intensity and linewidth for the (0.5,0) spectrum with just a local spin picture. Our results call for a more unified mechanism that reconciles these features and embraces the two types of excitations as having a common origin. 
 
\begin{center}
$\textbf{ACKNOWLEDGEMENT}$
\end{center}

The work at NIST is supported by the US Department of Commerce. SPINS and MACS utilized facilities supported in part by the NSF under Agreement No. DMR-0944772. 
The work at UT is supported by the U.S. DOE, BES, through DOE Grant No. DE-FG02-05ER46202.  The work at IOP is supported by the 
Chinese Academy of Science.

\end{document}